\documentclass[preprint,12pt]{elsarticle}
\usepackage{amssymb}
\usepackage{amsmath}
\usepackage[T1]{fontenc}
\usepackage{color}
\journal{Physica A}

\begin{document}
\begin{frontmatter}

\title{Dynamical phase transitions in  two-dimensional Brownian Matter}

\author[a]{Nathan O.  Silvano}
\affiliation[a]{
organization={Institute for Cross-Disciplinary Physics and Complex Systems (IFISC), CSIC-UIB}, 
 addressline={Ctra. Valldemossa km 7.5}, 
city={Palma de Mallorca},
postcode={07122}, 
 s%tate={RJ},
 country={Spain} }

\author[b]{Daniel G.  Barci}
\affiliation[b]{
organization={Departamento de F{\'\i}sica Te\'orica,
Universidade do Estado do Rio de Janeiro }, 
 addressline={Rua S\~ao Francisco Xavier 524}, 
city={Rio de Janeiro},
postcode={20550-013}, 
 state={RJ},
 country={Brazil} }

\begin{abstract}
We investigate collective behavior of  a system of two-dimensional interacting Brownian particles in the hydrodynamic regime.   By means of the Martin-Siggia-Rose-Jenssen-de Dominicis formalism,  we built up a generating functional for correlations functions.   In the continuum limit,  we uncover an exact symmetry under area-preserving diffeomorphism transformations that characterizes a liquid state. This symmetry leads to the conservation of local vorticity.  By computing the generating functional within the saddle-point plus Gaussian fluctuations approximation,  we reveal the emergence of a $U(1)$ gauge  symmetry that allows us to describe the dynamics of density fluctuations as a gauge theory.  We solve the corresponding equations of motion for short as well as long ranged interactions  showing up the presence of multiple dynamical regimes and associated dynamical phase transitions, even for pure repulsive interactions.  
\end{abstract}

\begin{highlights}
\item A path integral representation of interacting Brownian particles in the hydrodynamic regime is built. 
\item The system has an  exact symmetry under area-preserving diffeomorphism transformations (APD) that characterizes a liquid state.
\item Invariance under  APD implies the  conservation of local vorticity.
\item In the small fluctuations approximation,  APD shows up as a $U(1)$ gauge symmetry.
\item Purely repulsive long-ranged interactions leads to dynamical phase transitions and pattern formation.
\end{highlights}


\end{frontmatter}

%%%%%%%%%%%%%%%%%%%%%%%%%%%%%%%%%
\section{Introduction}
%%%%%%%%%%%%%%%%%%%%%%%%%%%%%%%%%
The study of Brownian matter,  {\em i.e.},  a large number of interacting Brownian particles, offers a rich variety of phenomena to explore.  It provides valuable insights into universal principles of statistical physics, such as the role of fluctuations and the importance of symmetry breaking in driving phase transitions \cite{VICSEK-2012}.

One of the most intriguing aspects of collective phenomena is pattern formation, where seemingly disordered systems give rise to structured arrangements and spatial organization. From the spontaneous emergence of intricate fractal patterns\cite{Vicsek-book-1991} to the development of complex self-assembled structures \cite{Grzybowski-2000}, Brownian matter serves as a platform for understanding the principles underlying pattern formation in nature\cite{ Cross-book-2009,Sevilla-2014, Dossetti-2015}.

Beyond its theoretical significance, the study of Brownian matter also has practical implications across diverse fields.
In materials science, insights into the self-assembly of colloidal particles \cite{Glotzer-2007, Manoharan-2015, wittkowskiDynamicalDensityFunctional2011} open pathways to designing novel materials with tailored properties, ranging from advanced ceramics to functional coatings.  In biophysics, Brownian motion has been used as a starting point for understanding the dynamics of biomolecules within cellular environments \cite{sheaBrownianMotionTheoretical1966, doi:10.1073/pnas.72.8.3111}. The temporal changes in cellular movement can be modeled by stochastic fluctuations \cite{jamali2020}, shedding light on key processes like protein folding and intracellular transport \cite{Alberts-2007,doi:10.1073/pnas.95.8.4288,dirienzoProbingShortrangeProtein2014}.

From a technical perspective, in addition to numerical simulations\cite{Sammuller-2021}, the mathematical methods used to tackle these systems are quite intricate. The hydrodynamic limit is frequently approached through the use of the Dean-Kawasaki (D-K) equation\cite{KAWASAKI-1994, DavidSDean_1996,illien-2024}. The central idea behind this method is similar to the Eulerian description of fluid dynamics, where the dynamical variables are densities and currents. The time evolution of particle density is governed by a non-trivial closed Langevin equation.

In this work, we adopt an alternative approach based on the Lagrangian description of fluid dynamics\cite{Rieutord-book-2015}, where the main dynamical variables are the positions of particles. Although both the Eulerian and Lagrangian descriptions are equivalent, their relationship is non-trivial
\cite{Jackiw2004}.

Using the Martin-Siggia-Rose-Janssen-De Dominicis (MSRJD) formalism \cite{MSR1973, Janssen-1976, deDominicis}, it is possible to exactly integrate the noise, leading to a generating functional for correlation functions. In this framework, the continuum limit can be easily taken, revealing a hidden symmetry of the system generated by area-preserving diffeomorphism transformations. A physical consequence of this symmetry is the conservation of local vorticity, connecting with the Kelvin circulation theorem in fluid mechanics \cite{Rieutord-book-2015,Jackiw2004}.  In the limit of weak density fluctuations, this symmetry manifests as a simple $U(1)$ gauge symmetry, providing an effective dual gauge theory for the dynamics of density fluctuations. In a previous paper \cite{Nathan-2024}, we applied this method to a system of active particles, focusing on orientational degrees of freedom. Here, we explore the consequences of these symmetries in a passive system with various types of microscopic interactions, particularly focusing on the dynamical regime and the conditions for dynamical phase transitions.

The primary result of this paper is the dual gauge theory represented by the generating functional of Eq. (\ref{eq:ZA}), with the effective action in Eq. (\ref{eq:SGauge}). This gauge theory resembles quantum electrodynamics (QED) in the imaginary-time path integral formalism. The emergent ``magnetic field" represents weak density fluctuations around a homogeneous background, while the emergent ``electric field" tracks the conservation of local vorticity, which, in the effective theory, appears as ``charge" conservation. The non-local ``magnetic susceptibility" is determined by microscopic two-body interactions.

We have investigated systems characterized by both short-range and long-range microscopic interactions. 
For repulsive short-range interactions, the stable steady state corresponds to a uniform density. Any small density fluctuation is dynamically suppressed and decays to zero. Conversely, attractive short-range interactions can lead to instabilities above a critical temperature, enabling the propagation of sound waves within the system.
Long-range interactions exhibit significantly richer and more complex behavior, even when purely repulsive. We have analyzed dipolar interactions\cite{Golden-2010}  and a class of soft-core interactions \cite{Delfau-2016}.  In both cases, we observed the emergence of dynamical phase transitions that demarcate boundaries between dissipative and dispersive regimes. Within these transition regions, the uniform density state becomes unstable, leading to the formation of spatial patterns.

The paper is structured as follows: In Section \ref{S:ABM}, we introduce the model and formalism. In Section \ref{S:Symmetries}, we present the central result of the paper, i.e., the emergence of a gauge symmetry and the development of the effective action for density fluctuations. In Section \ref{S:Dynamics}, we study the dynamics of density fluctuations, and in Sections \ref{S:LP} and \ref{S:NLP}, we discuss results on dynamical phase transitions for specific microscopic two-body interactions. Finally, in Section \ref{S:Discussions}, we present our conclusions, leaving some mathematical details of the calculations in Appendix \ref{App:MSRJD}.

%%%%%%%%%%%%%%%%%%%%%%%%%%%%%%%%%%
\section{Interacting two-dimensional Brownian particles}
\label{S:ABM}
%%%%%%%%%%%%%%%%%%%%%%%%%%%%%%%%%%
The simplest model for $N$ two-dimensional  interacting  Brownian particles is given by a  system of overdamped  Langevin equations, 
\begin{align}
\frac{d {\bf r}_i(t)}{dt}&=-\sum_{j\neq i} \boldsymbol{\nabla}U(|{\bf r}_i-{\bf r}_j|))+ \boldsymbol{\xi}^T_i(t) \; ,
\label{eq:Langevin}
\end{align}
with  white noise,  
\begin{align}
\langle \xi^T_{\alpha,i}\rangle &=0 \\
\langle \xi^T_{i,\alpha}(t)\xi^T_{j,\beta}(t')\rangle&=2k_B T\delta_{ij}\delta_{\alpha\beta}\delta(t-t') \; .
\end{align}
${\bf r}_i(t)$ with $i=1,\ldots,N$ are the two-dimensional position vectors  of each particle as a function of time.  
In these equations,  Latin indexes $i,j=1,\ldots, N$ label the particles, while the Greek indexes $\alpha,\beta=1,2$ are  Cartesian components in the plane $r_\alpha\equiv \{x_1,x_2\}$.     $U(|{\bf r}_i-{\bf r}_j|))$ is a pair potential between particles and  $k_B T$ is the thermal energy of the bath ($k_B$ is the Boltzmann constant). Along the paper,  we use bold characters for vector quantities.

We are interested in studding symmetries and collective behavior of this system.   For this,  a functional formalism is an appropriate tool.

%%%%%%%%%%%%%%%%%%%%%%%%%%%%%
\subsection{Functional formalism}
The path integral representation of a stochastic process is a powerful tool for calculating correlation and response functions~\cite{Lubensky2007, arenas2010}. This approach leverages the extensive machinery developed in quantum field theory, particularly non-perturbative techniques such as the dynamical renormalization group~\cite{Goldenfeld, Janssen-RG, Nathan-2023}. Additionally, this formalism highlights fundamental symmetries that provide insight into conserved quantities. Notably, non-trivial symmetries~\cite{arenas2012, Arenas2012-2} play a crucial role in governing the dynamical pathway toward equilibrium.  
In this section, we employ the Martin-Siggia-Rose-Janssen-de Dominicis (MSRJD) formalism~\cite{MSR1973, Janssen-1976, deDominicis} to construct the generating functional. Specifically, we follow the general methodology outlined in Ref.~\cite{Miguel2015}. Details of the formalism are provided in Appendix~\ref{App:MSRJD}.

The generating functional can be cast in the form,   
\begin{align}
Z[\boldsymbol{\eta}]=&\int \left(\prod_i{\cal D}{\bf r}_i(t)\right)
\label{eq:Z} \\
&\times    \exp\left\{-\frac{1}{2k_B T}S[{\bf r}_i]+\sum_i \int dt \;\boldsymbol{\eta}_i \cdot {\bf r}_i\right\} 
\nonumber 
\end{align}
where, as before,  $i=1,\ldots,N$ labels each particle.  
The action is given by 
\begin{align}
S= \sum_i \int  dt \left\{ \frac{1}{2} \left|\frac{d {\bf r}_i(t)}{dt}\right|^2 +{\cal V}({\bf r_i})\right\}
\label{eq:ST}
\end{align}
with the effective potential 
\begin{equation}
{\cal V}({\bf r_i})=\frac{1}{2} \left|\boldsymbol{\nabla}_{r_i}\tilde U\right|^2 
-k_BT \nabla_{r_i}^2 \tilde U
\label{eq:nu}
\end{equation}
in which we have defined 
\begin{equation}
\tilde U({\bf r}_i)=\sum_{j\neq i} U(|{\bf r}_i-{\bf r}_j|)  \;  .
\label{eq:Utilde}
\end{equation}

The  explicit form of the effective potential depends on the particular discretization scheme used to define the Wiener integral in the stochastic integration.  Equation (\ref{eq:nu}) is the expression for the Stratonovich prescription.  Notice that, since we are dealing with additive noise Langevin equations,   observables (noise mean values) do not depend on any specific discretization.  However, the path integral representation does depend on the prescription\cite{arenas2010} (please,  see appendix \ref{App:MSRJD} for details).

This formalism is an exact representation of the system of Langevin equations. Once $Z[\boldsymbol{\eta}]$ is known, we can compute any correlation function by functional differentiating
\begin{equation}
\left\langle r_{i_1}(t_1)\ldots r_{i_n}(t_n)\right\rangle_{\boldsymbol{\xi}}=\left. \frac{\delta^{(n)} Z[\boldsymbol{\eta}]}{\delta\eta_{i_n}(t_n)\ldots\delta\eta_{i_1}(t_1)}\right|_{{\boldsymbol{\eta}}=0}  \; .
\end{equation}

It is interesting to observe that the stochastic process given by the Langevin equations (\ref{eq:Langevin})  has the same path integral representation (Eqs. (\ref{eq:Z})-(\ref{eq:ST})) than a quantum particle system interacting with the potential ${\cal V}({\bf r})$ in the euclidean  imaginary time representation.   

%%%%%%%%%%%%%%%%%%%%
\section{Symmetries and the continuum limit}
\label{S:Symmetries}
%%%%%%%%%%%%%%
The continuum theory of perfect fluids, deduced from a microscopic description of particles obeying deterministic equations of motion (Newton's Laws), has been explicitly developed in Refs.  
\cite{Susskind-1991, Jackiw2004}. In our case, the interacting particles obey stochastic differential equations which, depending on the density, can lead to a liquid state.
  In this section,  we generalize the results of Ref. \cite{Jackiw2004} to the case of  stochastic dynamics.  To do this,  we perform a continuum limit at the level of the generating functional instead of the original stochastic differential equations. The advantage of this procedure is that,  since the noise has been already integrated,   we can work out the generating functional by means of  the usual tools of quantum field theory. 

Following Ref.  \cite{Jackiw2004},  consider a system with a large number of particles at finite mean density $\rho_0=N/A$, {\em i.e.}, the number of particles $N$ per unit area $A$ remains finite, even in the limit $N\to\infty, A\to\infty$.
  Under this assumption, we  can take the continuum limit.   For this,  we  simply promote the particle label  $i=1,\ldots,N$ to a continuum two-dimensional vector variable 
${\bf  y}\equiv (y_1,y_2)$, in such a way that  
\begin{align}
i &\to  {\bf y}
\label{eq:iy} \\
{\bf r}_i(t)& \to {\bf r}({\bf y}, t)   \; . 
\label{eq:riry}
\end{align}
The set of position vectors ${\bf r}_i$ with $i=1,\ldots,N$ turns out to be a two-dimensional vector field ${\bf r}({\bf y}, t)$.
Usually,  the coordinate ${\bf y}$ is fixed by demanding that it describes the actual particle position at initial time
\begin{equation}
{\bf r}({\bf y}, 0)={\bf y}\; , 
\label{eq:ic}
\end{equation}
thus,  the actual space dimension and the comoving coordinate (${\bf y}$) dimension coincide.

Sums over particles turn out to transform into integrals, 
\begin{align}
\sum_i&\to \rho_0\int d^2y \; .
\label{eq:sumi}
\end{align}
Then,  the action of   Eq. (\ref{eq:ST}) can be written in the continuum limit  as  
\begin{equation} 
 S=
  \rho_0 \int  dt d^2y \; \;  {\cal L}\left({\bf r}({\bf y}, t), \partial_t {\bf r}({\bf y}, t)\right)
\label{eq:SK}
\end{equation}
where $\partial_t$ is the time partial derivative.  The Lagrangian density is given by 
\begin{equation} 
 {\cal L}\left({\bf r}({\bf y}, t), \partial_t {\bf r}({\bf y}, t)\right)=
 \frac{1}{2}\left|\partial_t {\bf r}({\bf y}, t)\right|^2 + {\cal V}\left({\bf r}({\bf y}, t)\right)
\label{eq:L}
\end{equation}
with  the effective potential
\begin{align}
{\cal V}\left({\bf r}({\bf y}, t)\right)&=\frac{\rho_0^2}{2}  \int  d^2y' d^2y''  \nonumber \\
& \times \boldsymbol{\nabla} U({\bf r}({\bf y})-{\bf r}({\bf y}'))\cdot\boldsymbol{\nabla} U({\bf r}({\bf y})-{\bf r}({\bf y}''))
\nonumber \\
& -k_BT\rho_0
\int dt  d^2y' \nabla^2 U({\bf r}({\bf y})-{\bf r}({\bf y}')) \; .
\label{eq:SU}
\end{align}
There is a subtle assumption in Eq. (\ref{eq:SK}): we assume that not only the position ${\bf r}({\bf y},t)$, but also the velocity ${\bf v}({\bf y},t)=\partial_t{\bf r}({\bf y},t)$, is a smooth function of ${\bf y}$.
  If this is not the case,   the MSRJD formalism is not more appropriated,  and a more detailed formalism in terms of kinetic equation in phase space is mandatory\cite{Jackiw2004}. 

Thus,  Equation (\ref{eq:SK}) is the Lagrange formulation of a fluid,  described in terms of the physical position of the  particles ${\bf r}({\bf y},t)$.  
The classical equation of motion reads, 
\begin{equation}
\partial^2_t r_\alpha({\bf y},t)=\frac{\delta{\cal V}({\bf r})}{\delta r_\alpha({\bf y},t)} \; .
\label{eq:Classical}
\end{equation}
In general,   ${\cal V}({\bf r})$ is a complicated non-local function of the particle positions.  However, if the interactions are  reasonable short ranged,  it can be assumed that the potential ${\cal V}({\bf r})$ depends on the density at a point ${\bf r}$ and its derivatives: 
\begin{equation}
{\cal V}({\bf r}(y))\sim {\cal V}(\rho({\bf r}(y)),\nabla\rho({\bf r}(y)))
\end{equation}
 where $\rho({\bf r})$ is the particle density.    In this case,  Equation (\ref{eq:Classical}) leads to the Euler's  equation for a perfect fluid\cite{Jackiw2004,  Susskind-1991, Rieutord-book-2015}.  Here, we wont assume this property a priori.  Instead,  we will perform  a systematic low temperature  approximation of the generating functional providing a broader description of the system.  

%%%%%%%%%%%%%%%%%
\subsection{Area-preserving diffeomorphisms and vorticity}
\label{subsec:APD}
The system of  Langevin equations,  Eq. (\ref{eq:Langevin}),  has an obvious symmetry under  renumbering the particle labels $i$.  
In the  Lagrange continuum formulation of the system,  this symmetry shows up as an invariance under area preserving  diffeomorphisms.   Let us elaborate on this important concept.
 
There is huge freedom in making the transition  to the continuum limit.  For instance,  instead of the variable ${\bf y}$, given by equation  (\ref{eq:iy}),   we could choose another parametrization ${\bf y}'$, as for instance
\begin{align}
i &\to  {\bf y}={\bf f}({\bf y}')
\label{eq:ify} 
\end{align}
where ${\bf f}({\bf y})$ is an  invertible smooth vector function.
Under this transformation, equation (\ref{eq:riry}) transforms as, 
\begin{align}
{\bf r}_i(t)& \to {\bf r}({\bf y},t)={\bf r}({\bf f}({\bf y}'), t)\equiv {\bf r}'({\bf y}')  \; .
\label{eq:riry'}
\end{align}  
Therefore, each component of the position vector transforms as a scalar under reparametrizations
\begin{equation}
 r'_\alpha({\bf y}')=r_\alpha({\bf y}) \; .
 \label{eq:scalar}
\end{equation}   
with $\alpha=1,2$.

Moreover, from equation (\ref{eq:sumi}), we find  
\begin{align}
\sum_i&\to \rho_0\int d^2y=\rho_0\int d^2y'   \det\left(\frac{\partial y_i}{\partial y'_j} \right) 
\label{eq:sumii}
\end{align}
where the $\det\left(\partial y_i/\partial y'_j \right)$ is the Jacobian of the transformation of equation (\ref{eq:ify}).   As a consequence,  the action of the system,  equation (\ref{eq:SK}),  
is invariant under this reparametrization,  
\begin{equation}
S'[{\bf r}'({\bf y}', t)]=S[{\bf r}({\bf y}, t)]\; , 
\end{equation}
provided the Jacobian of the transformation  is  one,  {\em i.  e. }, 
\begin{equation}
J({\bf y}')\equiv\det\left(\frac{\partial  y_i}{\partial y'_j}\right)= \det\left(\frac{\partial f_i({\bf y}')}{\partial y'_j}\right)=1 \; .
\label{eq:Jacobian=1}
\end{equation}
In this case,  the transformation of Eq. (\ref{eq:ify}) is a mapping  between the planes $\{y_1,y_2\}\to \{y'_1,y'_2\}$ that preserves the area contained in any closed contour.
To see this,  consider for instance a domain $D\subset \{y_1,y_2\} $ enclosing an area $\Omega$.   The transformation of Eq. (\ref{eq:ify}) changes the domain  from $D\to D'\subset \{y'_1,y'_2\}$, enclosing a new area $\Omega'$.   Thus, 
\begin{equation}
\Omega=\int_{D} d^2y= \int_{D'} d^2y' \; J({\bf y}')  \; .
\end{equation}
In the last equality,  we have used equation  (\ref{eq:ify}) to transform ${\bf y} \to {\bf y}'$ and $D\to D'$.   Due to the fact that the Jacobian $J({\bf y}')=1$ (equation (\ref{eq:Jacobian=1})),  then, $\Omega'=\Omega$.

Therefore,   the action $S[{\bf r}({\bf y}, t)]$ has an {\em exact symmetry} given by its invariance under   general {\em area preserving diffeomorphisms } (APD). Moreover, since $r_\alpha({\bf y},t)$
is a scalar field under APD,  the functional measure is also invariant,  
\begin{equation}
{\cal D}{\bf r}({\bf y},t)={\cal D}{\bf r}'({\bf y}',t)
\end{equation}
implying the absence of anomalies.   Therefore,  an APD transformation is not only an invariance of the action by an exact symmetry of the system. 

This symmetry is a general consequence of the fluid state.  Indeed, it immediately implies   the vanishing of the static shear modulus\cite{Dubovsky2012,Nicolis2024, Matteo}.  The generalization of this symmetry for fluids in higher dimensions can be also worked out without difficulty\cite{Jackiw2004}.

No\"ether theorem associates a conserved quantity with any continuous symmetry of the action. To compute it,  it  is sufficient to consider only infinitesimal transformations.  Explicitly,   an infinitesimal  APD  can be written in the following way,  
\begin{equation}
y'_i=y_i+\epsilon_{ij}\frac{\partial \Lambda({\bf  y})}{\partial y_j} \; , 
\label{eq:APD}
\end{equation}
where $\epsilon_{ij}$ is the completely antisymmetric Levi-Civita tensor and  $\Lambda({\bf  y})$ is an arbitrary function satisfying  
\begin{equation}
\left(\frac{\partial}{\partial y_1}\frac{\partial}{\partial y_2}-\frac{\partial}{\partial y_2}\frac{\partial}{\partial y_1}\right)\Lambda({\bf  y})=0 \; . 
\label{eq:crossD}
\end{equation}
It is a simple matter to check from equations (\ref{eq:APD}) and (\ref{eq:crossD}), by direct computation,  that 
\begin{equation}
\det(\frac{\partial y'_i}{\partial y_j})=1+ O(\Lambda^2) \; .
\end{equation}
Thus,  equation (\ref{eq:APD}) represent an APD at linear order in $\Lambda$.

 From  equations (\ref{eq:scalar}) and (\ref{eq:APD}), the position vector transforms,  at linear order in $\Lambda$,  as 
\begin{align}
\delta r_\alpha({\bf y})&\equiv r'_\alpha({\bf y})-r_\alpha({\bf y})\nonumber \\
&=-\epsilon_{ij}\partial_i r_\alpha({\bf y})\partial_j \Lambda({\bf y}) +O\left(\Lambda^2\right) \; .
\label{eq:APDx}
\end{align} 
(From  now on,   Latin indexes $i,j=1,2$, label coordinates in the plane $\{y_1,y_2\}$  and the symbol  $\partial_j\equiv  \partial~/\partial y_j$.  In addition,  we continue using  Greek index $\alpha=1,2$  to label components of the  position vector  ${\bf r}$. ) 

The  first variation of the action  reads, 
\begin{align}
\delta S&=\int dt d^2y  
\left[\frac{\delta{\cal L} }{\delta r_\alpha({\bf y},t)}-\frac{d~}{dt}\left(\frac{\delta{\cal L} }{\delta \partial_t r_\alpha({\bf y}, t)}\right)\right]\delta r_\alpha({\bf y},t) \nonumber \\
&+\int dt d^2y \frac{d~}{dt}\left[\frac{\delta{\cal L} }{\delta \partial_t r_\alpha({\bf y},t)}  \delta r_\alpha({\bf y},t)\right] 
\label{eq:variation}
\end{align}
Imposing $\delta S=0$ for fixed initial and final configurations, we arrive to the Euler-Lagrange equation for the field ${\bf r}({\bf y}, t)$ given explicitly by equation (\ref{eq:Classical}).
For field configurations that satisfy equation (\ref{eq:Classical}),   the condition $\delta S=0$ implies (form the second line of Eq. (\ref{eq:variation}) the conserved quantity
\begin{equation}
Q(t)\equiv \int d^2y \frac{\delta {\cal L}}{\delta \partial_t r_\alpha({\bf y},t)} \delta r_\alpha({\bf y},t)
\end{equation}
where 
\begin{equation}
\frac{dQ(t)}{dt}=0.
\end{equation}
Using Eq. (\ref{eq:APDx}),  integrating by parts and asking for $Q$ to be a constant for any function $\Lambda({\bf y})$, we find
\begin{equation}
\frac{d~}{dt} \left[  \partial_j \left(\epsilon_{ij}\frac{d r_\alpha}{dt}\partial_i r_\alpha  \right)\right]=0 \; .
\end{equation}
Therefore,  due to the invariance of the system under  area preserving diffeomorphisms,  there is a  conserved quantity given by 
\begin{equation}
 \omega({\bf y})\equiv \partial_j {\cal J}_j
 \label{eq:vorticity}
\end{equation}   
where the current 
\begin{equation}
{\cal J}_j=\epsilon_{ij}\frac{d r_\alpha({\bf y},t)}{dt}\partial_i r_\alpha({\bf y},t)  \; .
\end{equation}

To see more clearly  the physical significance of $\omega({\bf y})$,  let us integrate this quantity in a region $\Omega$ bounded by the closed curve  $\partial \Omega$, 
\begin{equation}
\int_\Omega d^2y\;  \omega({\bf y})=\int_\Omega d^2y\; \partial_j{\cal J}_j({\bf y})=\oint_{\partial \Omega} {\cal J}_j \hat n_j d\ell 
\end{equation}
where, in the second equality,  we have used the divergence theorem.  $\hat n_j$ is a unit vector perpendicular to the curve $\partial \Omega$. 
Observing that $\epsilon_{ij} \hat n_j d\ell=dy_i$ over the curve $\partial \Omega$ we get
\begin{equation}
\int_\Omega d^2y\;  \omega(y)=\oint_{\partial \Omega}  \frac{d r_\alpha({\bf y},t)}{dt}\partial_i r_\alpha({\bf y},t) dy_i
\end{equation}
Changing variables using the mapping ${\bf r}={\bf r}(\bf y)$ and noting that $dr_\alpha=\partial_i r_\alpha({\bf y},t) dy_i$ we immediately obtain 
\begin{equation}
\int_\Omega d^2y\;  \omega(y)=\oint_{C[t]}  \frac{d r_\alpha}{dt} dr_\alpha
\label{eq:circulation}
\end{equation}
where $C[t]$ is the image of the closed curve $\partial\Omega$ in real position space,  produced by the mapping $C[t]= \bar{\bf r}[\partial\Omega, t]$,  where $\bar{\bf r}({\bf y},t)$ is a solution of the Euler-Lagrange equation  (\ref{eq:Classical}).
The right hand side of equation (\ref{eq:circulation}) is the circulation of the fluid velocity on the closed curve $C[t]$ that moves with the fluid stream.  The conservation of the circulation of the velocity is known in fluid mechanics as the {\em Kelvin circulation theorem}\cite{Rieutord-book-2015} and is applied to barotropic fluids subject solely to forces deriving from a potential.  
In this paper,  we are showing that the same result can be applied to overdamped Brownian particles interacting via  two-body potentials.  The deep reason behind this conservation is the invariance of the system under area preserving diffeomorphisms.   

For instance,  if   $\omega=0$,  the fluid has no circulation and this condition is kept by the dynamics.  In this sense,   $\omega({\bf y})$ can be interpreted as a local vorticity. 

%%%%%%%%%%%%%%%%%%%%%%%%%
\subsection{Saddle point plus Gaussian fluctuations: Emergent U(1) gauge symmetry}
In this section we evaluate the generating functional  in the saddle point approximation. 
The main idea is to write the vector position as
\begin{equation}
r_\alpha({\bf y},t)=r^0_\alpha({\bf y},t)+\delta r_\alpha({\bf y},t)
\end{equation} 
where $r^0_\alpha({\bf y},t)$ is a solutions of Eq. (\ref{eq:Classical}) and $\delta r_\alpha({\bf y},t)$ represent small fluctuations.  Then,  the generating functional is written as
\begin{align}
Z&=\exp\left\{-\frac{1}{2k_B T}S({\bf r}^0)\right\}\int  \left({\cal D}\delta{\bf r}\right)
\label{eq:Zsp} \\
&\times \exp\left\{-\int dt d^2y d^2y' \; \delta r_\alpha({\bf y},t)S_{\alpha\beta}^{(2)}(y-y')\delta r_\beta({\bf y}',t)\right\} 
 \nonumber
\end{align}
where the kernel
\begin{equation}
S_{\alpha\beta}^{(2)}(y-y')=  \left. 
\frac{\delta^2 S}{\delta r_\beta({\bf y}',t)\delta r_\alpha({\bf y},t) }
\right|_{{\bf r}={\bf r}^0}  \; .
\end{equation}
Linear terms in $\delta r_\alpha$ automatically cancel since ${\bf r}^0$ is an extreme of the action.  Moreover,    in equation (\ref{eq:Zsp}),  we have dropped out cubic and higher order terms in  $\delta r_\alpha$. 

The next step is to compute $r^0({\bf y}, t)$ and to conveniently parametrize fluctuations. 
The simplest solution of Eq. (\ref{eq:Classical}), which satisfies the initial condition of Eq. (\ref{eq:ic}),  is the static function
\begin{equation}
r^0_\alpha({\bf y},t)=y_\alpha\; .
\label{eq:r0}
\end{equation}
It is a simple matter to demonstrate that, for reasonable shorted ranged  potentials
\begin{equation}
\left. \frac{\delta{\cal V}({\bf r})}{\delta r_\alpha({\bf y},t)}\right|_{{\bf r}({\bf y},t)= {\bf y}}=0  \; .
\end{equation}
Fluctuations around this solutions are conveniently parametrized by an arbitrary vector field ${\bf A}({\bf y},t)$, 
\begin{equation}
\delta r_\alpha({\bf y},t)=\frac{1}{\rho_0}\epsilon_{\alpha\beta} A_\beta({\bf y}, t)
\label{eq:fluctuations}
\end{equation}
 in such a way that the position vector field is given by
\begin{equation}
r_\alpha({\bf y},t)=y_\alpha+\frac{1}{\rho_0}\epsilon_{\alpha\beta} A_\beta({\bf y}, t)
\label{eq:rA}
\end{equation}
The choosing of this parametrization resides in the fact that,  an infinitesimal APD,  given by 
Eq. (\ref{eq:APD}),  is now represented by  
\begin{equation}
A_i({\bf y})\to A_i({\bf y})+\rho_0 \partial_i \Lambda(\bf y) 
\label{eq:gauge}
\end{equation}
which is a usual  $U(1)$ gauge transformation.

Before proceeding to compute the generating functional,  let us look deeper into the physical interpretation of Eqs.  (\ref{eq:rA}) and (\ref{eq:gauge}).  The particle density  is  defined as
\begin{equation}
 \rho({\bf r}, t)=  \sum_i \delta^2\left({\bf r}-{\bf r}_i(t)\right) 
\label{eq:density}
\end{equation}
where ${\bf r}_i(t)$ is the position of the $i^{\rm th}$ particle as a function of time.  
In the continuum limit, it takes the form, 
 \begin{equation}
 \rho({\bf r},t)=  \rho_0 \int d^2y'\;   \delta^2\left({\bf r}-{\bf r}({\bf y}', t)\right) 
\label{eq:density-cont}
\end{equation}
By replacing  ${\bf r}({\bf y}',t)={\bf y}'$ in equation (\ref{eq:density-cont}), we trivially find $ \rho({\bf r},t)=  \rho_0$.  Thus,  the static solution of the equation of motion  (Eq. (\ref{eq:r0})) represents a static uniform density state $\rho_0$.  For more complex density configurations,  we need a careful treatment of Eq. (\ref{eq:density-cont}).    For instance,  by using the property of the $\delta$-function
\begin{equation}
\delta^2(g_\alpha({\bf y}'))=\frac{\delta^2({\bf y}'-{\bf y})}{|\det(\partial_i g_\alpha)|}
\end{equation}
where ${\bf y}$ is a single root of ${\bf g}({\bf y}')$,  ({\em i.e.},   $g_\alpha({\bf y})=0$) and choosing $g_\alpha({\bf y}')=r_\alpha-r_\alpha({\bf y}', t)$,   we find the density written in the plane $\{y_1,y_2\}$
\begin{equation}
\rho({\bf y},t)\equiv\rho({\bf r}({\bf y}, t)) = \rho_0\det\left(\frac{\partial y_i}{\partial r_\alpha}\right)
\label{eq:density-y}
\end{equation}
where the determinant is the Jacobian of the inverse mapping ${\bf y}\equiv {\bf y}({\bf r}) $.
Using the parametrization of Eq. (\ref{eq:rA}) and computing the Jacobian of Eq. (\ref{eq:density-y}) at linear order in ${\bf A}$ we find
\begin{equation}
\rho({\bf y},t)=\rho_0+ \boldsymbol{\nabla}\times {\bf A}({\bf y},t)+\ldots
\end{equation}
Therefore,  density fluctuations around a uniform background are parametrized by an emergent ``magnetic field''  $B=\boldsymbol{\nabla}\times {\bf A}$,  since
\begin{equation}
\delta\rho({\bf y},t)\equiv\rho({\bf y},t)-\rho_0= B({\bf y},t)
\end{equation}

From the point of view of symmetries, note that the  particle density $\rho({\bf r},t)$ (Eq. \ref{eq:density-cont}), similarly to the action $S[{\bf r}({\bf y},t)]$,   is exactly  invariant under APDs.   
This huge symmetry leads to a simpler $U(1)$ gauge symmetry when parametrized in terms of an emergent magnetic field.   However,  it is necessary to bare in mind that,  the $U(1)$ emergent gauge symmetry is appearing in the  small fluctuation regime, where  
$\delta\rho/\rho_0= B/ \rho_0<<1$.

Let us now go back to the computation of the generating functional of Eq. (\ref{eq:Zsp}) using the parametrization of Eq. (\ref{eq:rA}.)  First, we focus in the kinetic term of the action.
Using Eq. (\ref{eq:rA}) we immediately obtain
\begin{align} 
 S_K&= \frac{\rho_0}{2} \int  dt d^2y\left|\partial_t {\bf r}({\bf y}, t)\right|^2 \nonumber \\
  &=\frac{1}{2} \int  dt d^2y\left|\partial_t {\bf A}({\bf y}, t)\right|^2
\label{eq:SKA}
\end{align}
Moreover,  using Eqs. (\ref{eq:vorticity}) and  (\ref{eq:fluctuations}),   we can write the condition of zero vorticity as
\begin{equation}
\omega({\bf y},t)=\frac{1}{\rho_0} \boldsymbol{\nabla}\cdot \partial_t{\bf A}=0  \; .
\label{eq:omegaA}
\end{equation}
We can identify Eqs. (\ref{eq:SKA}) and (\ref{eq:omegaA}),  as the usual electric field action,  complemented with the Gauss law  in the temporal gauge $A_0=0$.     We can incorporate the field $A_0$ as a Lagrange multiplier, in order to   get the Gauss law (zero vorticity) as an equation of motion.  In this way, we can write $S_K$ in an explicit gauge invariant form, 
\begin{equation} 
 S_K=
  \frac{1}{2} \int  dt d^2y\left|{\bf E}({\bf y}, t)\right|^2  \; ,
\label{eq:SKE}
\end{equation}
where we have defined  the  emergent  ``electric field"  ${\bf E}=-\boldsymbol{\nabla} A_0-\partial_t {\bf A}$.   By  functional deriving $S_K$ with respect to $A_0$ we obtain the Gauss law $\boldsymbol{\nabla}\cdot {\bf E}=0$ in an explicit gauge invariant form, which means that the fluid has zero circulation.  

Terms containing two-body potentials  are given by 
\begin{equation}
S_{\cal V}=  \rho_0 \int  dt d^2y \; \;  {\cal V}\left({\bf r}({\bf y}, t)\right)
\end{equation}
Using Eq. (\ref{eq:rA}) and expanding in functional Taylor series up to second order in ${\bf A}$, 
\begin{align}
&{\cal V}\left({\bf r}({\bf y}, t)\right)={\cal V}\left({\bf y}\right)
+\frac{\epsilon_{\alpha\gamma}}{\rho_0} \int d^2y'    
\left.  
\frac{\delta{\cal V}}{\delta r_\alpha({\bf y}')}
\right|_{{\bf r}({\bf y})={\bf y}}\!\!\!\!\!\!\!\!
A_{\gamma}({\bf y}')
\nonumber 
\\
& +\frac{\epsilon_{\alpha\gamma}\epsilon_{\beta\sigma}}{2\rho_0^2} \int d^2y'd^2y''  A_{\gamma}({\bf y}')  
\left.  
\frac{\delta^2 {\cal V}}{\delta r_\alpha({\bf y}')\delta r_\beta({\bf y}'')}
\right|_{{\bf r}({\bf y})={\bf y}}\!\!\!\!\!\!\!\!
A_{\sigma}({\bf y}'')
\nonumber \\
&+\ldots
\end{align}

By explicitly computing the functional derivatives,   using Eq. (\ref{eq:SU}), 
we find for the complete  action $S=S_K+S_{\cal V}$ 
\begin{align}
S=
 \frac{1}{2} &\int  dt d^2y\left|{\bf E}({\bf y}, t)\right|^2 
   \label{eq:SGauge}  \\
  +&\frac{1}{2}\int dt d^2 y d^2y' B({\bf y},t) V({\bf y}-{\bf y}') B({\bf y}',t)
 \nonumber
\end{align}
in which
\begin{align}
V({\bf y}-{\bf y}')&=  
\frac{\rho_0}{2}\int d^2z \boldsymbol{\nabla}_z  U({\bf y}-{\bf z})\cdot \boldsymbol{\nabla}_z  U({\bf z}-{\bf y}')
\nonumber \\
&-k_BT\; \nabla^2_y U({\bf y}-{\bf y}')   \; .
\label{eq:V} 
\end{align}
The effective action,  written in this form,  is explicitly Gauge invariant.   The functional integration measure is written in term of  fluctuations as ${\cal D}{\delta\bf r}\to  {\cal D}{\bf A} $. However,  since configurations of the field ${\bf A}$ connected by the gauge transformation of Eq. (\ref{eq:gauge}) represent the same physical state,  we need to restrict the integration  over classes of gauge configurations as it is usually done in any gauge theory\cite{Fradkin-Book-2021}
\begin{equation}
{\cal D}{\bf r}\to  {\cal D}{\bf A} \delta(G_\Lambda({\bf A})) \det[\delta_\Lambda G_\Lambda]
\end{equation}
where $G_\Lambda({\bf A})$ is a gauge fixing function with parameter $\Lambda$ and the Jacobian 
$ \det[\delta_\Lambda G_\Lambda]$,  usually known as the Faddeev-Popov determinant,  guarantee gauge invariance of the integration measure. 
Therefore,   the generating functional for correlation functions is finally given by 
\begin{equation}
Z=\int {\cal D}{\bf A} \delta(G_\Lambda({\bf A})) \det[\delta_\Lambda G_\Lambda]\;  e^{-\frac{1}{2k_B T} S[{\bf A}]}
\label{eq:ZA}
\end{equation}
where $S[{\bf A}]$ is given by Eq. (\ref{eq:SGauge}).

Equations  (\ref{eq:SGauge}) and (\ref{eq:ZA}) are the main result of the paper.  
The system of interacting Brownian particles, in the small density fluctuation approximation,  is completely equivalent to an $U(1)$ gauge theory that resembles a  `` quantum electromagnetic theory".
Let us emphasize again that this {\em emergent} $U(1)$ symmetry is not an exact symmetry of the hole system.  It is a  manifestation of the exact invariance under area preserving diffeomorphisms in the limit of small fluctuations around a constant density.   

In this framework, the dynamics of small density fluctuations are effectively captured by a gauge theory. The gauge field ${\bf A}({\bf y}, t)$ encodes the particle positions in the Lagrangian description of the fluid. Since density is a physical observable, it must be a gauge-invariant quantity, represented as  
$B = {\bf \nabla} \times {\bf A}$.   
Another fundamental observable is the vorticity, which relates to the fluid's angular momentum and is expressed as   ${\bf \nabla} \cdot {\bf E}$.   Thus, the local vorticity of the fluid acts as a dual electric charge.  
Analogous to the role of gauge fields (photons) mediating interactions between charges in electromagnetic theory, in this dual framework, density fluctuations mediate interactions between vortices. By mapping the problem onto an effective gauge theory with emergent fields ${\bf E}$ and $B$, this approach bridges statistical mechanics, fluid dynamics, and field theory, offering a unified perspective. This framework provides a powerful tool to investigate phenomena such as dynamical phase transitions and pattern formation.  
In this study, we focus on the case of zero vorticity. Under this condition, the dynamics of density fluctuations simplify to the propagation of "electromagnetic" fields in the absence of charge distributions. The effects of charges (vortices) and their dynamics will be addressed in a future work.
  
%%%%%%%%%%%%%%%%%%%
\section{Dynamics  of density fluctuations }
\label{S:Dynamics}
%%%%%%%%%%%%%%%%%%%
The dynamics of small density fluctuations are given by  the equations of motion
\begin{align}
&\frac{\delta S}{\delta A_0({\bf y},t)}=0 \; , 
\label{eq:deltaSA0}\\
&\frac{\delta S}{\delta A_i({\bf y},t)}=0 \; .
\label{eq:deltaSAi}
\end{align}
where $S$ is the action of Eq. (\ref{eq:SGauge}). 

Equation (\ref{eq:deltaSA0}) is simply the Gauss law
\begin{equation}
\boldsymbol{\nabla}\cdot {\bf E}=0  \; . 
\label{eq:GaussLaw}
\end{equation}
On the other hand,  Faraday's law is automatically satisfied due to gauge invariance, 
\begin{align}
&\boldsymbol{\nabla}\times {\bf E}+\partial_t B=0 \; .
\label{eq:Faraday}
\end{align}
Finally,  Eq. (\ref{eq:deltaSAi}) leads to the equation 
\begin{align}
\partial_t E_i({\bf y}, t)+ \rho_0 \epsilon_{ij}\int d^2y' V({\bf y}-{\bf y}') \partial_j B({\bf y}', t)=0  \; .
\label{eq:Ampere}
\end{align}
Eqs. (\ref{eq:GaussLaw}), (\ref{eq:Faraday}) and (\ref{eq:Ampere}) complete the set of integro-differential equations that determines the dynamics of the ``electromagnetic" fields  $\{{\bf E}({\bf y},t), B({\bf y},t)\}$.
Since we are interesting in the dynamics of density fluctuations ( $\delta\rho({\bf y},t)=B({\bf y},t))$,  we can write an equation only in terms of the magnetic field $B$. For this, we take the curl of equation (\ref{eq:Ampere}) and use the Faraday law,  equation (\ref{eq:Faraday}), obtaining, 
\begin{align}
\partial_t^2 B({\bf y},t)+\rho_0\int d^2y'  V({\bf y}-{\bf y}') \nabla^2 B({\bf y}', t)=0   \; .
\end{align}
This equation can be cast in terms of the original potential $U({\bf y}-{\bf y}')$ by using Eq. (\ref{eq:V}). We find, 
\begin{align}
&\partial_t^2 B({\bf y},t)- \int d^2y' \left\lbrace \rho_0 k_B T\; U({\bf y}-{\bf y}')\right.  \\
& \left.+\frac{\rho^2_0}{2}\int d^2z  \; U({\bf y}-{\bf z})U({\bf z}-{\bf y}')\right\}  \nabla^4 B({\bf y}', t)=0  \; .
\nonumber 
\end{align}
Since this is a linear equation, we can write it in Fourier space $\{\omega, {\bf k}\}$.  By  introducing the Fourier transform
\begin{equation}
B({\bf y}, t)=\int \frac{d\omega}{2\pi} \frac{d^2k}{(2\pi)^2} \tilde B(\omega,{\bf k}) e^{-i\omega t+i{\bf k}\cdot{\bf y} }
\label{eq:Byt}
\end{equation} 
we obtain
\begin{align}
&\left(\omega^2 + k_B T\rho_0 \left\lbrace  \tilde U({\bf k})+\frac{\rho_0}{2k_BT}  \tilde U^2({\bf k})\right\}k^4\right)  \tilde B({\bf k},\omega)=0 \; . 
\label{eq:Bkw}
\end{align}
where 
\begin{equation}
 \tilde U({\bf k})=\int d^2 y \;  U({\bf y})  e^{-i{\bf k}\cdot{\bf y} } \; .
\end{equation}
Since Eq. (\ref{eq:Bkw}) is an homogeneous linear equation,  solutions with $\tilde B({\bf k},\omega)\neq 0$
only exist if the following dispersion relation is satisfied
\begin{equation}
\omega^2=- k_B T\rho_0 \left\lbrace  \tilde U({\bf k})+\frac{\rho_0}{2k_BT}  \tilde U^2({\bf k})\right\}k^4  \; .
\label{eq:dispersion}
\end{equation}
The general solution for small density fluctuations in the Brownian medium is given by 
equation (\ref{eq:Byt}) with the dispersion given by equation (\ref{eq:dispersion}).

The propagation of waves in the Brownian medium essentially depends on the behavior of the Fourier transform of the two-body potential $\tilde U ({\bf k})$.  Form Eq. (\ref{eq:dispersion}) it is clear that 
if $\tilde U ({\bf k})>0$  for all values of $k$,   then the dispersion is purely imaginary
\begin{equation}
\omega=\pm i  \sqrt{ k_B T\rho_0} \left\lbrace  \tilde U({\bf k})+\frac{\rho_0}{2k_BT}  \tilde U^2({\bf k})\right\}^{1/2} k^2
\label{eq:I-dispersion}
\end{equation}
In this case,  any weak perturbation of the constant density $\delta\rho({\bf r},t)=B({\bf r},t)$ will be dumped to zero with the dispersion of Eq. (\ref{eq:I-dispersion}). It is important to note that both, positive and negative imaginary parts are solutions of the linear equation. However, the growing solution rapidly gets out of the approximation of weak perturbations and it will be controlled by non-linear terms  that we are ignoring.  

It is interesting to note that, in the  low temperature regime $k_B T/\rho_0U<< 1 $,  Eq. (\ref{eq:I-dispersion}) reduces to
\begin{equation}
\omega\sim i  \frac{\rho_0}{\sqrt{2}} k^2  U(\bf k)\; .
\end{equation}
This result exactly coincides with the one computed using the D-K formulation\cite{Delfau-2016}.

Moreover, if $\tilde U ({\bf k})<0$ for some values of $k$,  then several interesting possibilities arises.   Sound waves with wave vector  ${\bf k}$  can be propagated if $\tilde U ({\bf k})<-2k_B T/\rho_0$. Thus, sound propagation is a thermal property that may appears when thermal fluctuations overwhelm the typical interaction energy.  In addition,  for   $\tilde U ({\bf k})=-2k_B T/\rho_0$,   static pattern formation is possible.   

In the next section we will discuss some interesting examples of  specific two-body potentials.

%%%%%%%%%%%%%%%%%%%%%%%%%
\section{Short-range potentials}
\label{S:LP}
%%%%%%%%%%%%%%%%%%%%%%%
The simplest possible two-body  short-range interaction is given by the  potential  
\begin{equation}
U({\bf y}-{\bf y}')=U_0\;  \delta^2\left({\bf y}-{\bf y}'\right)
\end{equation}
where the constant $U_0$ measure the  intensity of the potential.  $U_0>0$ produces local repulsion between particles while local attraction is modeled  by  $U_0<0$.   The Fourier transform is simply
$\tilde U({\bf k})=U_0$.  Replacing this value into Eq.  (\ref{eq:dispersion})  we find 
\begin{equation}
\omega=\pm i \frac{ \rho_0|U_0|}{\sqrt{2}}  \sqrt{ 1+\frac{2k_BT} {\rho_0 U_0}}\; k^2
\label{eq:dispersion-local}
\end{equation}
We see that for repulsive short-range potentials,  $U_0>0$, the system does not allow sound propagation. Any density fluctuation is damped to zero with a quadratic dispersion. 
However,  for  attractive potentials $U_0<0$ the situation changes. 
For small temperatures $k_B T<< \rho_0 |U_0|$,  collective excitations are overdamped quadratic modes, 
 \begin{equation}
\omega=\pm i  \rho_0|U_0| \; k^2 \; .
\label{eq:dispersion-localKT}
\end{equation}
Interestingly, if the thermal energy overwhelms the interaction energy $k_B T>> \rho_0 |U_0|$,   the system support sound waves with quadratic dispersion
\begin{equation}
\omega=\pm \sqrt{\frac{k_BT}{\rho_0|U_0|}} \; k^2
\end{equation}
It seems to be a dynamical phase transitions at a critical temperature 
\begin{equation}
k_B T_c=\frac{\rho_0 |U_0|}{2}
\end{equation}
where for $T>T_c$ the system of Brownian particles behaves as a medium where sound waves   propagate with quadratic dispersion.  Below this temperature,  any density fluctuation is damped.

%%%%%%%%%%%%%%%%%%%%%%%%%%%%
\section{Long-range potentials}
%%%%%%%%%%%%%%%%%%%%%%%%%%%%
\label{S:NLP}
Long-range  pair potentials are much more interesting than the short-range ones.  As we have previously shown, local repulsive  interactions do not allow sound propagation and they do not produce any non-homogeneous structure. In other words, the homogeneous constant density is stable under small perturbations. This situation changes for long-range potentials.  In the following, we show two examples of purely repulsive long-range potentials that give rise to dynamical phase transitions as well as pattern formation. 

%%%%%%%%%%%%%%%%%%%%%%%
\subsection{Two-dimensional  dipolar interaction}
\label{S:dipolar}
Consider  electric dipoles oriented in the $\hat z$ direction (perpendicular to the plane ${\bf r}$\cite{Golden-2010}.  The interaction between any pair of particles is completely repulsive,  given by the potential
\begin{equation}
U({\bf r})=\frac{\mu^2}{r^3}
\label{eq:Ur-dipolar}
\end{equation}
where $\mu^2$ is the electric dipole strength. 
The Fourier transform is given by,  
 \begin{equation}
\tilde U({\bf k})=-2\pi \mu^2 k
\label{eq:Uk-dipolar}
\end{equation}
Note that $\tilde U({\bf k})<0$ for all values of $k$. 
Replacing Eq. (\ref{eq:Uk-dipolar}) into Eq. (\ref{eq:dispersion}) we find
\begin{equation}
\omega^2= k_BT \rho_0 2\pi\mu^2\;  k^5\left\{1-\frac{2\pi \mu^2\rho_0}{2k_BT} k\right\}
\label{eq:dispertion-dipolar}
\end{equation}
Clearly,  there is a scale,   given by  wave vector
\begin{equation}
k_0=\frac{2k_BT}{2\pi \mu^2\rho_0}
\label{eq:k0-dipole}
\end{equation} 
that separates two very different dynamical behavior,  {\em i.e.}, 
\begin{equation}
\omega=
\left\{
\begin{array}{lcl}
\pm\sqrt{4\pi k_BT \rho_0 \mu^2} \; k^{5/2} & \mbox{for} &  k<<k_0 \\
&  & \\
\pm i  2\pi\mu^2\rho_0 \;  k^{3} & \mbox{for} &  k>>k_0 \\
\end{array}
\right.
\label{eq:disp-dipolar-limits}
\end{equation}
The system can propagate sound waves with wave-vectors $k<<k_0$. The dispersion in this case is $\omega\sim k^{5/2}$.
Conversely,  density fluctuations  with wave-vectors $k>>k_0$ are overdamped with cubic dispersion $\omega\sim i k^3$. 

Near $k\sim k_0$ we have the following non-analytic dispersion
\begin{equation}
\omega\sim  \pm 2k_B T k_0^{3/2} \sqrt{k_0- k}  \; .
\label{eq:k-k0}
\end{equation}
At exactly $k=k_0$,  $\omega=0$.  This means that the system have the tendency to form static patterns with wave vector of modulus $k_0$.   

From Eqs.~(\ref{eq:Byt}) and (\ref{eq:k-k0}), we can infer different possible patterns, such as linear stripes:  
\begin{equation}
\rho({\bf y}) - \rho_0 = B({\bf y}) \sim \cos\left({\bf k_0} \cdot {\bf y}\right),
\end{equation}  
or even circular defects:  
\begin{equation}
\rho(y) - \rho_0 = B(y) \sim \cos\left(k_0 y\right).
\end{equation}  
It is worth noting that in two dimensions, fluctuations generally destroy these patterns, leading to the formation of more complex structures. For example, a nematic order can emerge, where rotational symmetry is spontaneously broken while translational symmetry remains intact~\cite{BaSt2007-1, BaSt2009}.  
However, the dispersion relation in Eq.~(\ref{eq:k-k0}) is non-analytic, distinguishing this theory from conventional models of pattern formation, such as the Brazovskii model~\cite{Br1975} or the Swift-Hohenberg model~\cite{SwHo1977}, where the dispersion is analytic, \(\omega \sim (k^2 - k_0^2)\). The square-root singularity in Eq.~(\ref{eq:k-k0}) introduces a sudden change in the dynamics, with the branch point at \(k = k_0\) resembling a non-Hermitian "exceptional point" singularity~\cite{Rui-2020}.  
Therefore, the actual mechanism of pattern formation in Brownian matter requires further investigation.

%%%%%%%%%%%%%%%%%%
\subsection{Soft core interactions}
\label{S:GEP}
A convenient way of modeling soft-core repulsive interaction is a family of generalized exponential potentials given by \cite{Delfau-2016}
\begin{equation}
U_m({\bf r})=\epsilon \exp\left\{-\left(\frac{r}{\sigma}\right)^m\right\} \; ,
\label{eq:GEP}
\end{equation}
where $m$ is the order of the potential,   $\sigma$ is the effective range,  in the sense that for $r>>\sigma$ the potential is essentially zero and   $\epsilon$ is the highest energy,  reached at $r=0$.  

The Fourier transform of equation (\ref{eq:GEP}) is given by
\begin{align}
\tilde U_m(\sigma k) &=\epsilon \sigma^2 \int_0^\infty dx\;  x e^{-x^m} J_0(\sigma k x) \; ,
\label{eq:GEPk}
\end{align}
where $J_0(\sigma k x)$ is the Bessel function  of the first kind of  order zero. 
It is well known that for $m\leq 2$,  $U_m>0$. The simplest example is the Gaussian potential, $m=2$, whose Fourier transform is
\begin{align}
\tilde U_2(\sigma k)&=\frac{\epsilon \sigma^2}{2} e^{-\left(\frac{\sigma k}{2}\right)^2}  \; .
\end{align}
In this case, the system of Brownian particle is unable to propagate sound waves, since the dispersion relation 
of equation (\ref{eq:dispersion}) is purely imaginary for all values of $k$.  Then,  any density fluctuations is dumped to zero and the only static solution is $\omega=k=0$. Thus,  the homogeneous solution 
$\rho({\bf r},t)=\rho_0$ is stable under small fluctuations. 

However, for $m>2$,   the Fourier transform of equation (\ref{eq:GEP}) can take negative values for some values of the wavevector $k$. This opens the posibility of sound propagation as well as pattern formation. 
Let us analyze the case of $m=4$.   In this case, the integral in equation (\ref{eq:GEPk}) can be computed exactly,   obtaining
\begin{align}
\tilde U_4(\sigma k)&=\epsilon \sigma^2\left[\frac{\sqrt{\pi }}{4} \, _0F_2\left(\frac{1}{2},1;\left(\frac{\sigma k}{4}\right)^4\right) \right.  \nonumber \\
&\left.  ~~~~ -\left(\frac{\sigma k}{4}\right)^2 \, _0F_2\left(\frac{3}{2},\frac{3}{2};\left(\frac{\sigma k}{4}\right)^4\right)\right] \; , 
\label{eq:U4k}
\end{align}
where $_0F_2(b,c;z)$ is the generalized hypergeometric function of type $(0,2)$.

\begin{figure}[h]
\begin{center}
 \includegraphics[width=\columnwidth]{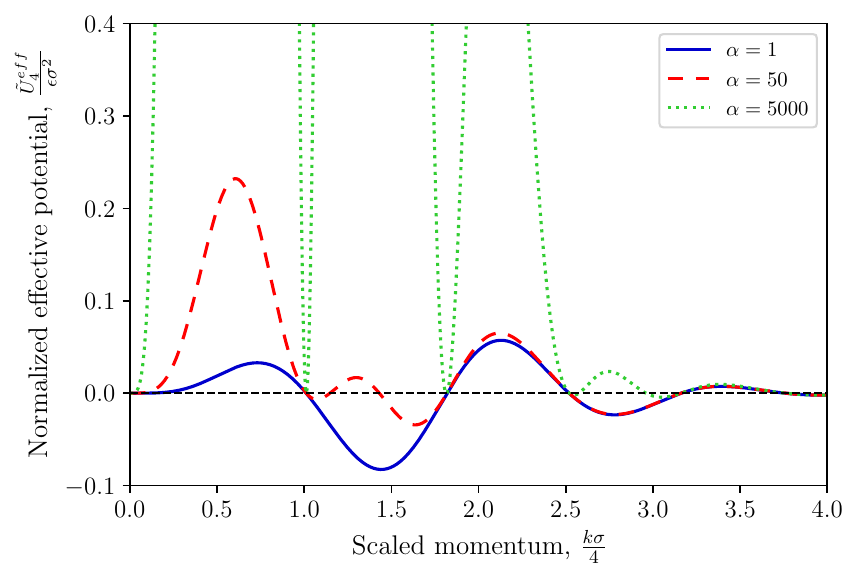}
\end{center}
\caption{Plot of effective potential $\tilde U_4^{\rm eff}$ (Eq.\,(\ref{eq:Ueff4})) normalized in units of $\epsilon\sigma^2$ as a function of scaled momentum $k\sigma/4$ for different temperature ratios $\alpha =\sigma^2\rho_0\epsilon/2k_BT $. The continuous (blue) line corresponds with $\alpha=1$,   in the dashed (red) line $\alpha=50$.  The extremely low temperature regime  $\alpha=5000$ is depicted in the dotted (green) line.}
\label{fig:Ueff4}
\end{figure}

The  dispersion relation,  equation (\ref{eq:dispersion}) takes the form
\begin{equation}
\omega^2=-2 k_B T\rho_0  \tilde U_{4}^{\rm eff}(k) \; . 
\label{eq:dispersionU4}
\end{equation}
where the effective potential $\tilde U_{4}^{\rm eff}(k)$ is given by
\begin{equation}
\tilde U_{4}^{\rm eff}(k)=\left\lbrace  \tilde U_4(k)+\frac{\rho_0}{2k_BT}  \tilde U_4^2(k)\right\}k^4 \; .
\label{eq:Ueff4}
\end{equation}
In the low temperature regime $k_B T<< \sigma^2\rho_0 \epsilon$,   the dispersion is dominated by 
$\tilde U_{4}^{\rm eff}(k)\sim \tilde U_4^2(k)\geq 0$.   In this case, the system only admits overdamped modes,  or static patterns with wavevectors $k_j$,  where $U_4^{\rm eff}(k_j)=0$. 
However,   for $k_B T\gtrsim\sigma^2\rho_0 \epsilon$,   the linear term proportional to $U_4(k)$ dominates the spectrum and there are regions in the wave vector space where $\tilde U_{4}^{\rm eff}(k)<0$.  In figure (\ref{fig:Ueff4}) we show $U_4^{\rm eff}(k)$ for different values of the parameter $\alpha =\sigma^2\rho_0\epsilon/2k_BT$.
%%%%%%%%%%%%%%%
%%%%%%%%%%%%%%
We clearly see the appearance of ``bands"  in which   $U_4^{\rm eff}(k)<0$ for a broad range of temperatures regime,  allowing sound propagation in these regions. 
Also, there are dynamical phase transition points $k_i$,  with $i=1,2,\dots$,  where there are static solutions,  representing periodic pattern formation at wavevectors  where  $U_4^{\rm eff}(k_i)=0$.  
Near these singularities, the dispersion relation is 
\begin{equation}
\!\omega_j\sim 
\left\{
\begin{array}{lcl}
\pm\sqrt{(-1)^{j}\sigma\left(k_j-k\right)} & \mbox{for} & k_B T\gtrsim\sigma^2\rho_0 \epsilon  \\
& & \\
\pm i\left|k_j-k\right| &\mbox{for} & k_B T<<\sigma^2\rho_0 \epsilon 
\end{array}
\right.
\end{equation} 
where $|k-k_j|\sigma <<1 $. 
We observe dynamical phase transitions in the high temperature regime for special values of momenta.  Very near the singularities, the dispersion relation has the same global properties that the previous dipolar case, {\em i.e.},  square root singularities. 
In this way,  the propagation of a pulse with a wide range of wave-vectors is a mixture of dispersive and dissipative regimes depending on the  wave-vector spectrum.
It is worth to mention that pattern formation of Brownian particles with repulsive soft core interactions has been observed in numerical simulations of the original Langevin equations\cite{Golden-2010}.

%%%%%%%%%%%%%%%%%%%%%%%%%%%%
\section{Discussions and conclusions}
\label{S:Discussions}
%%%%%%%%%%%%%%%%%%%%%%%%%%%%
We investigated the collective properties of a system of two-dimensional interacting Brownian particles in the hydrodynamic regime. Using the MSRJD formalism, we developed a generating functional for correlation functions, allowing us to take the continuum limit for a large number of interacting particles. This approach offers an alternative to the Dean-Kawasaki (D-K) equation. The key differences between the two methods are as follows: While the D-K formalism employs a Langevin equation for particle density,  resulting in stochastic multiplicative noise dynamics, our approach uses a path integral formulation in terms of particle positions. While the D-K formalism parallels the Eulerian description of fluid dynamics, with density and current as the primary variables, our method aligns with the Lagrangian description.  

We have shown that, in the Lagrangian framework, a system of interacting Brownian particles exhibits an exact symmetry in the continuum limit; specifically, an invariance under area-preserving diffeomorphisms. One significant physical consequence of this symmetry is the vanishing of the static shear modulus, indicating that the system exists in a fluid state.  
Unlike solids, where deformation is constrained by the underlying lattice, liquids lack such rigidity. The APD symmetry reflects this flexibility, allowing fluid elements to flow freely while conserving their density distribution.  
The conservation of vorticity arises naturally from the APD symmetry through Noether's theorem, which links symmetries to conserved quantities. Specifically, APD symmetry ensures that the circulation of velocity along a closed curve remains invariant over time. Physically, this implies that the fluid's rotational patterns—such as whorls, vortices, or eddies—are stable features of the dynamics, unaffected by the internal rearrangements of the fluid.  
This property is a hallmark of ideal liquid behavior, where it is known as Kelvin’s circulation theorem~\cite{Rieutord-book-2015}.

By computing the generating functional using the saddle-point approximation and Gaussian fluctuations, we demonstrated that APD invariance manifests as a $U(1)$ gauge symmetry, resulting in an effective gauge action for fluctuations around a homogeneous background. The primary contribution of this paper is captured in equations (\ref{eq:SGauge}) and (\ref{eq:ZA}). In this dual gauge theory,  ``electric charge" corresponds to vorticity in the original particle system, while the ``magnetic field" represents density fluctuations.

Using this framework, we described the dynamics of density perturbations through a set of deterministic integro-differential equations analogous to Maxwell's equations. Gauss's law arises from vorticity conservation, and Faraday’s law is inherently satisfied due to gauge invariance. These equations are fundamentally dictated by symmetry. The system is completed by the dynamic equation (\ref{eq:Ampere}), which incorporates information about the microscopic two-body potential in the original Brownian particle system.

The general solutions to these equations depend on the interplay between two-body interactions and thermal fluctuations. In the extremely low-temperature regime, density fluctuations are damped, suggesting that the homogeneous solution is stable, with any weak perturbation decaying to zero within a finite timescale that depends on the perturbation’s wave vector. This result aligns with the solution of the D-K equation under the same conditions. However, when thermal energy becomes comparable to or exceeds the typical interaction energy, more interesting phenomena, such as sound propagation or static pattern formation, can occur.

We solved the dynamical equations for various microscopic potentials. For a repulsive short-range potential, the only solution is overdamped behavior with quadratic dispersion. Conversely, for a local attractive potential, a critical temperature exists above which wave propagation occurs in a dispersive medium, also with quadratic dispersion. Thus, the critical temperature marks the boundary between two dynamic phases: in one, collective fluctuations are overdamped, and in the other, wave propagation is possible.

We also explored long-range repulsive interactions. In the case of a two-dimensional dipolar interaction $U(r) \sim 1/r^3 $, we identified a dynamical phase transition, even for a purely repulsive potential, between more exotic phases. The competition between thermal fluctuations and microscopic interactions defines a wave vector scale  $k_0$. At low temperatures or for wave vectors $k > k_0$, fluctuations are overdamped with a dispersion $\omega \sim i k^3$.  However, long-wavelength modes with $k < k_0$ can propagate without dissipation, with a dispersion $\omega \sim k^{5/2}$.  In the case of soft-core repulsive potentials, we found even richer behavior. For exponentially decaying potentials (faster than Gaussian), we identified dynamical phase transitions at specific wave vectors, separating regions of overdamped and dispersive wave propagation. In this scenario, bands for wave propagation exist, with their width and structure dependent on temperature.

Interestingly, in the cases involving long-range potentials, near or at the dynamical phase transitions, the system may develop pattern formation. Such structures were observed in Langevin simulations of soft-core potentials \cite{Delfau-2016} similar to the examples studied here. Moreover, we found that the dispersion relation near the transition is non-analytic, differing from the conventional Swift-Hohenberg-Brazovskii models of pattern formation. For the cases we examined, the square-root singularity resembles the structure of non-Hermitian exceptional point singularities, suggesting the possibility of non-trivial topological structures in the dynamics of collective modes.

In summary, we have provided an analytical method to study the dynamics of density fluctuations in two-dimensional Brownian particles in the hydrodynamic regime. This method reveals an emergent symmetry governing the system’s primary conservation laws. As emphasized, this gauge symmetry arises in the weak fluctuation regime. It would be valuable to explore the role of non-linearities, extending beyond this approximation. A systematic perturbative expansion that preserves the exact invariance under area-preserving diffeomorphisms could be a fruitful approach. We hope to report on the non-linear response of interacting Brownian systems and their potential topological structures in future work.

%%%%%%%%%%%%%%%%%%%%%%%%%%%%%%
\section*{Acknowledgments}
The Brazilian agencies, {\em Funda\c c\~ao Carlos Chagas Filho de Amparo \`a Pesquisa do Estado do Rio
de Janeiro} (FAPERJ), {\em Conselho Nacional de Desenvolvimento Cient\'\i
fico e Tecnol\'ogico} (CNPq) and {\em Coordena\c c\~ao  de Aperfei\c coamento de Pessoal de N\'\i vel Superior}  (CAPES) - Finance Code 001,  are acknowledged  for partial financial support.
During this work,  N.O.S.  was partially supported by a Doctoral Fellowship by CAPES and it is currently  partially supported by the María de Maeztu project CEX2021-001164-M funded by the  MICIU/AEI/10.13039/501100011033.
%%%%%%%%%%%%%%%%%%%%%%%%%%%%%%%%%%%%%%%%%%%

\appendix
\section{Generating functional for  Brownian particles}
\label{App:MSRJD}

In this appendix, we review the MSRJD method that leads to the action of equation (\ref{eq:ST}). The procedure essentially follows reference \cite{Miguel2015}.

Consider the following system of stochastic equations:
\begin{align}
\frac{d {\bf r}_i(t)}{dt}&=-\sum_{j\neq i} \boldsymbol{\nabla}_{r_i}U(|{\bf r}_i-{\bf r}_j|))+\boldsymbol{\xi}_i(t)
\label{app:EqT}
\end{align}
where $i=1,\ldots,N$. We use bold letters to indicate two-dimensional vector quantities, and when necessary, we use Greek indices $\alpha, \beta=1,2$ to represent the components of two-dimensional vectors.

Equation (\ref{app:EqT}) is a set of overdamped Langevin equations for $N$ particles with positions ${\bf r}_1,\ldots,{\bf r}_N$. The particles interact via pairwise forces given by:
\begin{equation}
{\bf F}_i(\{{\bf r}\}) \equiv \boldsymbol{\nabla}_{r_i} \tilde U_i(\{{\bf r}\}) \equiv  \boldsymbol{\nabla}_{r_i}\sum_{j\neq i} U(|{\bf r}_i-{\bf r}_j|)).
\end{equation}
Here, ${\bf F}_i(\{{\bf r}\})$ is the force exerted on the $i^{\rm th}$ particle by the other $N-1$ particles via the potential $\tilde U_i(\{{\bf r}\})$, where $\{{\bf r}\}$ represents the set of positions ${\bf r}_1, \ldots, {\bf r}_N$ for all particles.

The vector white noise $\boldsymbol{\xi}_i(t)$ is defined by the following correlation functions:
\begin{align}
\langle \xi_{\alpha,i}\rangle &=0 \\
\langle \xi_{i,\alpha}(t)\xi_{j,\beta}(t')\rangle&=2k_B T\delta_{ij}\delta_{\alpha\beta}\delta(t-t')
\end{align}
where $k_B T$ is the diffusion constant, identified with the temperature of the environment.

In the following, we construct the generating functional for the correlation functions of equation (\ref{app:EqT}).

The generating functional is given by:
\begin{align}
Z[\boldsymbol{\eta}] &= \int \left(\prod_i{\cal D}{\bf r}_i(t)\right)\left\langle \delta\left({\bf O}\right) 
\det\left[
\frac{\delta {\bf O}}{\delta {\bf r}}  \right] 
\right\rangle_{{\bf \xi}} \nonumber \\
&\times  \exp\{ \int dt \left(\boldsymbol{\eta}_i \cdot {\bf r}_i\right)\}
\label{app:Z}
\end{align}
where $\boldsymbol{ \eta}_i$ are source terms used to compute correlation functions, and the brackets $\langle \ldots \rangle_\xi$ represent the stochastic expectation value. In equation (\ref{app:Z}), we have introduced the vector function:
\begin{align}
O^\alpha_{i}(t) &= \frac{d r^\alpha_i(t)}{dt} + \nabla_{r_i}^\alpha \tilde U_i - \xi^{\alpha}_{i}(t),
\end{align} 
and the operator in the determinant is given by:
\begin{align}
\frac{\delta O^\alpha_{i}(t)}{\delta r_j^\beta(t')}&=\left\{\delta_{ij}\delta^{\alpha\beta}\frac{d~}{dt}+ \nabla_{r_j^\beta}\nabla_{r_i^\alpha} \tilde U_i\right\}\delta(t-t').
\end{align}

The goal of this formalism is to compute the stochastic expectation value exactly, allowing us to express the result solely in terms of the particle trajectories ${\bf r}_i(t)$. To achieve this, we first exponentiate the delta function using auxiliary vectors ${\bf A}_i$:
\begin{align}
\delta\left({\bf O}\right)&=\int  \left(\prod_i{\cal D}{\bf A}_{i}(t)\right) e^{i\int dt\sum_i  A^\alpha_{i} O^\alpha_{i}}
\end{align} 
We also exponentiate the determinant using a set of independent vector Grassmann variables $\{ \boldsymbol{\bar\psi}_{i},\boldsymbol{\psi}_{i}\}$:
\begin{align}
&\det\left[
\frac{\delta {\bf O}}{\delta {\bf r}}
\right] = \int  \left(\prod_i{\cal D}\boldsymbol{\bar \psi}_{i}{\cal D}\boldsymbol{\psi}_{i}\right) \\
&\times\exp\left\{ \int dtdt' \sum_{ij}   \bar\psi_{i}^\alpha(t) \frac{\delta O^\alpha_{i}(t)}{\delta r_j^\beta(t')} \psi_{j}^\beta(t') \right\} \nonumber
\end{align}
With these steps, the noise enters the exponential linearly, allowing it to be integrated exactly. After collecting all the noise terms, we obtain:
\begin{align}
 \left\langle e^{-i\int dt\sum_i A_{i}^\alpha \xi_{i}^\alpha}  \right\rangle_{\boldsymbol{\xi}} = e^{-\int dt\sum_i  \left\{  k_B T\left|{\bf A}_{i} \right|^2 \right\}}
\end{align}

Next, we integrate over the Grassmann variables. Collecting all terms in the Grassmann fields yields:
%\begin{widetext}
\begin{align}
&\int  \left(\prod_i{\cal D}\boldsymbol{\bar \psi}_{i}{\cal D}\boldsymbol{\psi}_{i}\right)
 \exp\left\{\int dt \sum_i  \boldsymbol{\bar \psi}_{i}\cdot \frac{d\boldsymbol{\psi}_{i}}{dt}+ \sum_{ij}\left( \bar \psi^\alpha_{i} \psi^\beta_{j} \nabla_{r_j^\beta}\nabla_{r_i^\alpha} \tilde U_i  \right)\right\} \nonumber \\
 &=\exp\left\{  \int dt \sum_i  \frac{1}{2} \nabla^2_{r_i}\tilde U_i   \right\}
\end{align}
%\end{widetext}
The last result follows from the relation:
\begin{align}
\left\langle  \bar \psi^\alpha_{i}(t)  \psi^\beta_{j}(t')  \right\rangle&= \delta_{ij}\delta^{\alpha\beta} G_R(t-t'),
\end{align}
where $G_R(t-t')$ is the retarded Green's function of the operator $d/dt$, and $G_R(0)=1/2$ corresponds to the Stratonovich stochastic prescription.

After integrating over the noise and Grassmann variables, we find:
\begin{equation}
Z=\int \left(\prod_i{\cal D}{\bf r}_i(t){\cal D}{\bf A}_i(t)\right) \exp\left\{-\frac{1}{2k_BT}S\right\}
\end{equation}
with:
\begin{align}
S&= 2k_BT\int dt  \left\{ \sum_i k_B T\left|{\bf A}_{i}\right|^2 +i A_{i}^\alpha\left(\frac{d r^\alpha_i(t)}{dt}+ \nabla_{r_i^\alpha} \tilde U_i  \right) \right.\nonumber \\
&\left. - \frac{1}{2} \sum_{j\neq i}\nabla^2_{r_i}U \right\} \; .
\label{ap:SA}
\end{align}
The last step is to integrate over ${\bf A}_i$. This is a Gaussian integral and can be performed straightforwardly.

Collecting all terms together, the action in terms of ${\bf r}_i(t)$ can be written as:
\begin{align}
S&=\frac{1}{2}\int  dt \sum_i \left\{\frac{d {\bf r}_i(t)}{dt}+ \boldsymbol{\nabla}_{r_i} \tilde U_i  \right\}^2 \nonumber \\
&-k_BT\int dt\sum_i \nabla^2_{r_i} \tilde U(\{ {\bf r}\}) \; .
\label{app:ST}
\end{align}
Equation (\ref{app:ST}) is the Onsager–Machlup action corresponding to the stochastic equation (\ref{app:EqT}). The last term represents the Jacobian of the transformation from $\boldsymbol{\xi}$ to ${\bf r}$, in the Stratonovich prescription.

To rewrite equation (\ref{app:ST}) in a Lagrangian form, we expand the square in the first line and integrate by parts the cross term. This gives:
\begin{align}
S&= \int  dt \sum_i \left\{ \frac{1}{2} \left|\frac{d {\bf r}_i(t)}{dt}\right|^2 +\frac{1}{2} \left|\boldsymbol{\nabla}_{r_i}\tilde U\right|^2 
-k_B T \nabla_{r_i}^2 \tilde U\right\} \nonumber \\
& +\sum_j \left(\tilde U_j(t_f)-\tilde U_j(t_i)\right).
\label{app:STlagrangian}
\end{align}
The first line of this equation corresponds to equation (\ref{eq:ST}) in the main text. The second line is a constant term arising from the integration of total time derivatives. Although this term is important for calculating some equilibrium properties, it does not affect fluctuations given by the correlation functions of ${\bf r}_i$.

%\bibliographystyle{elsarticle-num} 
%\bibliography{stochastic-0124,References.bib}

\end{document}